\documentclass[preprint,12pt]{revtex4-1}

\begin{document}

\title{Surname statistics - Crossing the boundary between disciplines
\\Comment on ``Surname distribution in
population genetics and in statistical physics'' by Rossi}
\author{Seung Ki Baek}
\email{seungki@kias.re.kr}
\affiliation{School of Physics, Korea Institute for Advanced Study, Seoul
130-722, Korea}
\author{Beom Jun Kim}
\email{beomjun@skku.edu}
\affiliation{BK21 Physics Research Division and Department of Physics,
Sungkyunkwan University, Suwon 440-746, Korea}
\maketitle

In East Asia, the public used to be denoted by an idiom whose literal
English translation means `a hundred surnames'.
This expression is very accurate in Korea, where more than 99\% of the
entire population are covered by the most common hundred surnames (among three
hundreds in total). Even though there are about 4,000 surnames in China, the
most common hundred among them still comprise 85\% of the Chinese
population. In the United States, on the other hand, this expression
loses the meaning, because a hundred surnames can represent only one fifth of
the population at most. All these indicate that the surname distributions
are exceedingly heterogeneous and the degree of heterogeneity also varies to a
great extent across countries. Such differences may well be traced back to
anthropological explanations, but one should note at the same time that
the dynamics of surnames lends itself to an elegant mathematical formulation
known as the branching process, which has been a useful stochastic model in
physics and biology.
Unfortunately, few have been able to overcome the barriers between
the disciplines, which someone might attribute to a split between the
\emph{two cultures}~\cite{snow}. Now, Rossi's \cite{rossi} review fills the
gap and provides a helpful guide to explore this important subject.

After providing valuable literature review on the history of the development of 
surname studies, the author comes to the question:
What determines the different degrees of heterogeneity in surname
statistics? 
Rossi~\cite{rossi} devotes the last few sections to this issue
and presents a renormalization-group (RG) approach in section 10. The RG
formalism was developed in the seventies to deal with diverging
characteristic scales at a continuous phase transition. From the RG
viewpoint, if a system with coupling strength $K$ is rescaled by a factor of
$b$, the free energy per site transforms as $F(K) = b^{-1} F[K'(K)] +
g(K)$, where $K'$ is a renormalized coupling at the new scale and $g(K)$
represents non-singular coarse-grained contribution from smaller scales.
Comparing this to the recurrence relations for $n_t(z)$ and
$\eta_t (z)$ in section 10, one should however note that
the extra terms $\theta(z)$ and $\alpha N(0)z$ do not result from
coarse-graining but represent independent mechanisms to introduce new
surnames. How such input depends on the size of the population is
actually a \emph{relevant} field operator that determines the degree of
heterogeneity in surname statistics.

Even the origin of heterogeneity still remains as an intriguing
question, especially when considered as a manifestation of Zipf's
law~\cite{zipf,kims}. A surname distribution arises from stochasticity
subject to rules of the branching process, so one may think of it as the
most random outcome under certain constraints.
Note that the focus is not on a specific dynamics but rather on
\emph{constraints} out of it, and it offers a possibility that many details
of different processes may turn out to be irrelevant after all. Such a
viewpoint explains the ubiquity of Zipf's law and puts
the problem of surname distributions into a more general context.


\begin{thebibliography}{4}%
\makeatletter
\providecommand \@ifxundefined [1]{%
 \@ifx{#1\undefined}
}%
\providecommand \@ifnum [1]{%
 \ifnum #1\expandafter \@firstoftwo
 \else \expandafter \@secondoftwo
 \fi
}%
\providecommand \@ifx [1]{%
 \ifx #1\expandafter \@firstoftwo
 \else \expandafter \@secondoftwo
 \fi
}%
\providecommand \natexlab [1]{#1}%
\providecommand \enquote  [1]{``#1''}%
\providecommand \bibnamefont  [1]{#1}%
\providecommand \bibfnamefont [1]{#1}%
\providecommand \citenamefont [1]{#1}%
\providecommand \href@noop [0]{\@secondoftwo}%
\providecommand \href [0]{\begingroup \@sanitize@url \@href}%
\providecommand \@href[1]{\@@startlink{#1}\@@href}%
\providecommand \@@href[1]{\endgroup#1\@@endlink}%
\providecommand \@sanitize@url [0]{\catcode `\\12\catcode `\$12\catcode
  `\&12\catcode `\#12\catcode `\^12\catcode `\_12\catcode `\%12\relax}%
\providecommand \@@startlink[1]{}%
\providecommand \@@endlink[0]{}%
\providecommand \url  [0]{\begingroup\@sanitize@url \@url }%
\providecommand \@url [1]{\endgroup\@href {#1}{\urlprefix }}%
\providecommand \urlprefix  [0]{URL }%
\providecommand \Eprint [0]{\href }%
\providecommand \doibase [0]{http://dx.doi.org/}%
\providecommand \selectlanguage [0]{\@gobble}%
\providecommand \bibinfo  [0]{\@secondoftwo}%
\providecommand \bibfield  [0]{\@secondoftwo}%
\providecommand \translation [1]{[#1]}%
\providecommand \BibitemOpen [0]{}%
\providecommand \bibitemStop [0]{}%
\providecommand \bibitemNoStop [0]{.\EOS\space}%
\providecommand \EOS [0]{\spacefactor3000\relax}%
\providecommand \BibitemShut  [1]{\csname bibitem#1\endcsname}%
\let\auto@bib@innerbib\@empty
\bibitem [{\citenamefont {Snow}(1959)}]{snow}%
  \BibitemOpen
  \bibfield  {author} {\bibinfo {author} {\bibfnamefont {C.~P.}\ \bibnamefont
  {Snow}},\ }\href@noop {} {\emph {\bibinfo {title} {The Two Cultures}}}\
  (\bibinfo  {publisher} {Cambridge University Press},\ \bibinfo {address}
  {London},\ \bibinfo {year} {1959})\BibitemShut {NoStop}%
\bibitem [{\citenamefont {Rossi}(2013)}]{rossi}%
  \BibitemOpen
  \bibfield  {author} {\bibinfo {author} {\bibfnamefont {P.}~\bibnamefont
  {Rossi}},\ }\href@noop {} {\bibfield  {journal} {\bibinfo  {journal} {Physics
  of Life Reviews}\ }\textbf {\bibinfo {volume} {10}},\ \bibinfo {pages} {395}
  (\bibinfo {year} {2013})}\BibitemShut {NoStop}%
\bibitem [{\citenamefont {Baek}\ \emph
  {et~al.}(2011{\natexlab{a}})\citenamefont {Baek}, \citenamefont
  {Bernhardsson},\ and\ \citenamefont {Minnhagen}}]{zipf}%
  \BibitemOpen
  \bibfield  {author} {\bibinfo {author} {\bibfnamefont {S.~K.}\ \bibnamefont
  {Baek}}, \bibinfo {author} {\bibfnamefont {S.}~\bibnamefont {Bernhardsson}},
  \ and\ \bibinfo {author} {\bibfnamefont {P.}~\bibnamefont {Minnhagen}},\
  }\href@noop {} {\bibfield  {journal} {\bibinfo  {journal} {New J. Phys.}\
  }\textbf {\bibinfo {volume} {13}},\ \bibinfo {pages} {043004} (\bibinfo
  {year} {2011}{\natexlab{a}})}\BibitemShut {NoStop}%
\bibitem [{\citenamefont {Baek}\ \emph
  {et~al.}(2011{\natexlab{b}})\citenamefont {Baek}, \citenamefont {Minnhagen},\
  and\ \citenamefont {Kim}}]{kims}%
  \BibitemOpen
  \bibfield  {author} {\bibinfo {author} {\bibfnamefont {S.~K.}\ \bibnamefont
  {Baek}}, \bibinfo {author} {\bibfnamefont {P.}~\bibnamefont {Minnhagen}}, \
  and\ \bibinfo {author} {\bibfnamefont {B.~J.}\ \bibnamefont {Kim}},\
  }\href@noop {} {\bibfield  {journal} {\bibinfo  {journal} {New J. Phys.}\
  }\textbf {\bibinfo {volume} {13}},\ \bibinfo {pages} {073036} (\bibinfo
  {year} {2011}{\natexlab{b}})}\BibitemShut {NoStop}%
\end{thebibliography}
%
\end{document}